\documentclass[12pt]{article}
  
\usepackage{latexsym,amsmath,amssymb,theorem,epsfig,graphics}
\usepackage[usenames]{color} 
\usepackage{graphicx,color}
\usepackage{fancyvrb, pstricks}

\DeclareFontFamily{OT1}{rsfs10}{}
\DeclareFontShape{OT1}{rsfs10}{m}{n}{ <-> rsfs10 }{}
\DeclareMathAlphabet{\mathscript}{OT1}{rsfs10}{m}{n}

\newcommand{\beq}{\begin{equation}}
\newcommand{\eeq}{\end{equation}}

\newcommand{\bea}{\begin{eqnarray}}
\newcommand{\eea}{\end{eqnarray}}
\newcommand{\ba}{\begin{array}}
\newcommand{\ea}{\end{array}}

\newcommand{\msfb}[1]{\texttt{#1}} 
\newcommand{\msf}[1]{\text{\sf{#1}}} 
\newcommand{\bs}[1]{#1} 

\definecolor{SVcol}{rgb}{0.8, 0.8, 0.8}

\newcommand{\cbox}[1]{{\colorbox{SVcol}{{#1}}}}
\newcommand{\eref}[1]{(\ref{#1})}

\def\Tb{\bar{T}}
\def\Sb{\bar{S}}
\def\math{{\em Mathematica}}
\def\sing{{\em Singular}}
\def\sv{STRINGVACUA}
\definecolor{MathRed}{rgb}{0.5,0.0,0.0}

\begin{document}

\begin{titlepage}
  \begin{flushright}
	arXiv:0801.1508 [hep-th]
  \end{flushright}
  \vspace*{\stretch{1}}
  \begin{center}
     \Huge STRINGVACUA\\ \vspace{0.3cm}
   \large A \math\ Package for Studying Vacuum \\ Configurations in String Phenomenology
  \end{center}
  \vspace{0.5cm}
  \begin{center}
    \begin{minipage}{\textwidth}
      \begin{center}
        {\large
	James Gray${}^{1,2*}$,
        Yang-Hui He${}^{2,3\sharp}$,
        Anton Ilderton${}^{4\flat}$ and
        Andr\'e Lukas${}^{2\dagger}$
	}
      \end{center}
    \end{minipage}
  \end{center}
  \vspace*{1mm}
  \begin{center}
    \begin{minipage}{\textwidth}
      \begin{center}
        ${}^1$Institut d'Astrophysique de Paris and APC, 
        Universit\'e de Paris 7,\\
        98 bis, Bd.~Arago 75014, Paris, France\\[0.2cm]
        ${}^2$Rudolf Peierls Centre for Theoretical Physics, 
        University of Oxford,\\ 
        1 Keble Road, Oxford OX1 3NP, UK\\[0.2cm]
	${}^3$Merton College, Oxford, OX1 4JD and
	  Mathematical Institute, Oxford University,\\
	  24-29 St.~Giles', Oxford OX1 3LB, UK\\[0.2cm]
        ${}^4$School of Mathematics and Statistics, University of Plymouth, \\
        Drake Circus, Plymouth PL4 8AA, UK
      \end{center}
    \end{minipage}
  \end{center}
  \vspace*{\stretch{1}}
  \begin{abstract}
    \normalsize 
We give a simple tutorial introduction to the \math\ package \sv, which
is designed to find vacua of string-derived or inspired four-dimensional $N=1$ supergravities.
The package uses powerful algebro-geometric methods, as implemented in the free 
computer algebra system \sing, but requires no knowledge
of the mathematics upon which it is based. A series of easy-to-use \math\ modules are provided which can
be used both in string theory and in more general applications requiring fast polynomial computations. The 
use of these modules is illustrated throughout with simple examples.
  \end{abstract}
  \vspace*{\stretch{5}}
  \begin{minipage}{\textwidth}
    \underline{\hspace{5cm}}
    \\
    \footnotesize
    ${}^*$email: j.gray1@physics.ox.ac.uk \\
    ${}^\sharp$email: hey@maths.ox.ac.uk\\
    ${}^\flat$email: abilderton@plymouth.ac.uk\\
    ${}^\dagger$email: lukas@physics.ox.ac.uk 
  \end{minipage}
\end{titlepage}

\tableofcontents

\section{Introduction}
The systematic investigation of the turning points of a given potential
is one of the most fundamental problems in physics. In particular,
much of the current effort in making string theory reproduce the real
world is concentrated in the study of effective theories with potentials capable of
stabilising all of the moduli. The theoretical framework for deriving the
potential given the particular scenario of compactification -- be it
heterotic or type II, with or without fluxes -- has been considerably
developed. However, the analysis of the resulting effective theory, due to the
typically complicated forms of the (super-) potential, is often
computationally insurmountable.

In \cite{Gray:2006gn,Gray:2007yq} a novel, algorithmic approach to
this problem was presented. The key realisation of these papers was
that the typical potential is a ratio of, albeit complicated, polynomials in the
field variables (in the non-perturbative cases, dummy variables may be used as substitutions for
exponential terms, recovering polynomial expressions). The most efficient method of studying
systems of polynomials is {\it computational algebraic geometry}.

Recent advances in computer algebra, in particular the advent of
such packages as {\it Macaulay 2} \cite{m2} and \sing\ \cite{sing}, have rendered the subject
of computational algebraic geometry an increasingly significant tool in string theory
\cite{Anderson:2007nc,Gray:2006jb,Distler:2005hi}. 
As outlined in \cite{Gray:2006gn},
the starting point for our analysis is a (string-derived or inspired)
${\cal N}=1$ supergravity theory with complex chiral superfields $\phi_i$,
given in terms of  the associated K\"ahler potential $K\equiv K(\phi,\bar{\phi} )$ and
superpotential $W\equiv W(\phi )$. The scalar potential $V$ of such a theory is
given by the well-known formula,
\begin{equation}
\label{sugrav}
 	V=\exp (K/M^2)\left(K^{ij}F_i\bar{F}_j-\frac{3}{M^2}|W|^2\right)\, .
\end{equation}
Here $F_i\equiv\partial_iW+M^{-2}\partial_iKW$ are the F-terms, $K_{ij}=\frac{\partial^2
K}{\partial\phi_i\partial\bar\phi_j}$ is the K\"ahler metric, $K^{ij}$
its inverse and $M$ is the Planck mass. Our primary interest is in
analysing the vacuum properties of this potential. 

One tool which computational algebraic geometry provides us with to help in this analysis is the so-called saturation expansion \cite{Gray:2006gn}. This method takes the complicated polynomial equations describing the extrema of the potential of a system, and breaks them up into a series of simpler systems, each describing one particular locus of turning points. Further algorithmic methods then make it possible to single out extrema of interest and to extract the physical properties of the associated vacua. Another tool, called elimination, provides constraints on
the parameters of a model which must be satisfied for vacua of given types to exist.

The computer package \math\ has become an
indispensable tool in modern theoretical physics and most researchers
are familiar with its syntax. The algebraic geometry system \sing\ is
perhaps less known to the community interested in effective
supergravity theories. With this in mind, we have written a package
for \math, called \sv , which externally calls \sing\ where
appropriate while sustaining the user in the comfortable environment
of \math.  A range of \math\ modules, from low to high level, provide
both ease of use in string theory calculations as well as
flexibility for a wider range of applications.  In particular, the tools
described in the previous paragraph are implemented in module form - allowing the 
user to access them with a few simple \math\ commands.

The methods of \sv\ are of course far from restricted in use to problems in string phenomenology. Many of the modules will be of use in any problem involving polynomial computations. Another \math\ package which makes use of \sing\ is \cite{singm}.  In the ensuing pages we shall walk the reader through a quick and  illustrative tutorial of how to use the basic commands of the package. The appendix contains a glossary of terms.

Throughout this tutorial text in {\sf Serif font} indicates a function or object defined in \math, while commands to be entered by the user into a \math\ notebook are \colorbox{SVcol}{\texttt{typeset as so}}.

%
%
\section{Getting started: a simple supergravity model}
\label{start}

\subsection{A quick first look: {\sf SimpleMinimise}}
Installation of the package is easy. Instructions, along with the download of the package itself, can be found at the following URL.
\bea
\nonumber
\textnormal{http://www-thphys.physics.ox.ac.uk/user/Stringvacua}
\eea

In this tutorial, we describe use of the package in a Unix environment, focussing on the \math\ notebook interface although the package also works with the text interface or running in emacs. The only significant differences from a Windows environment are during installation, for further details 
see the URL given above. Once you have the package installed start up X11, if it is not already running, and then to load the package simply evaluate (note the single open quote!)\\
\newline
\cbox{\texttt{<<Stringvacua}\`{}}\\
\newline
A package header and links to examples and the browser help pages will be
displayed, and the package is ready for calculations. For illustration,
let us start with a very simple example based on one modulus $T=t+i\tau$, with K\"ahler potential and superpotential given by
\begin{equation}
\label{model1}
	K=-3\log (T+\bar{T})\, ,\quad W=a+bT+cT^2\; ,
\end{equation}
respectively. Here, $a$, $b$ and $c$ are (real) parameters.

We will first study the vacuum structure of this theory using the {\sf SimpleMinimise} command, which is designed to provide an elementary illustration of the package's use and to familiarise the user with its output. Evaluating the command
\begin{Verbatim}
SimpleMinimise[]
\end{Verbatim}
will result in the appearance of a dialogue box saying {\sf List of complex fields? (example: \{T\})}. In our case our complex field is indeed T, thus we simply type \cbox{\texttt{\{T\}}} and click OK. The next two dialogue boxes ask for the real and imaginary parts of the complex fields. If we want these to be given by $T=t+i \tau$ then we enter {\colorbox{SVcol}{{\texttt{\{t\}}}}} followed by {\colorbox{SVcol}{{\texttt{\{$\backslash$[Tau]\}}}}} 
(using the \math\ syntax). We are then prompted to provide the K\"ahler potential, \cbox{\texttt{-3 Log[T+Conjugate[T]]}}, followed by the superpotential, which for our example is \cbox{\texttt{a + b T + c T$^\wedge$2}} and then the parameters in the model, \cbox{\texttt{\{a,b,c\}}}. Finally (in the unix release) the package asks how much time it is allowed to spend on each calculation that it will perform in \sing. Let us set this to 20 (seconds) for this example, by entering \cbox{\texttt{20}}.

\begin{figure}[ht]\centering\leavevmode\epsfysize=7cm 

\epsfbox{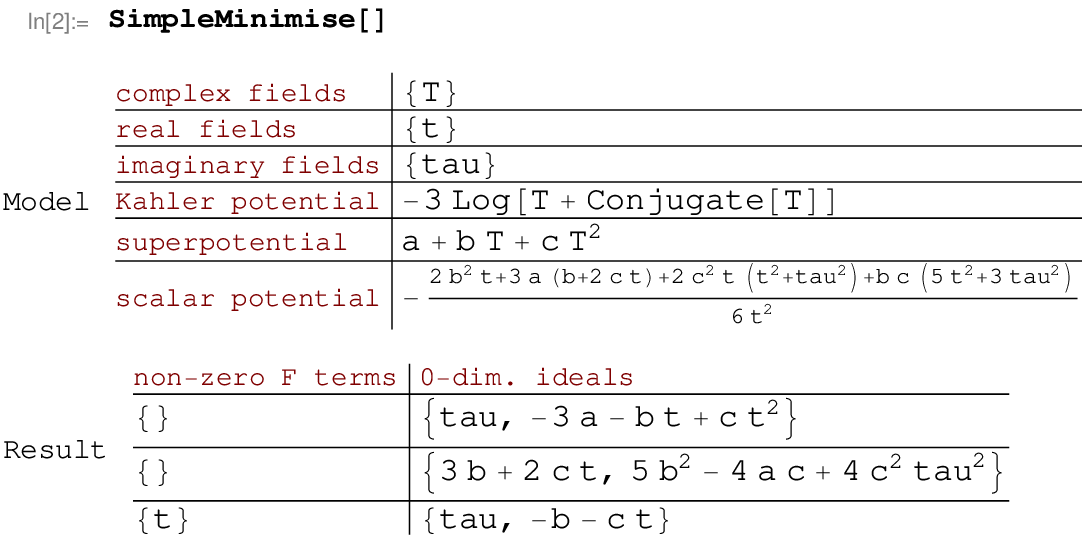}\\

\caption{An example of the output of {\sf SimpleMinimise}.}

\label{figure1}

\end{figure}

This data having been entered, the package then works out everything else automatically. An xterm will appear on your screen (by default in the upper right corner) to which \sing\ writes its protocol. The details of this protocol should not concern us for now: we will simply use its presence as an indication that something is ``happening" in our calculation. The final output looks as pictured in figure \ref{figure1}.

Let us discuss the interpretation of this result. Clearly the first few rows of the first table, labelled ``Model", simply repeat the data defining the model which we have input. The last entry in this table is the scalar potential of the system, which the package has calculated using the standard ${\cal N}=1$ four dimensional supergravity formula \eqref{sugrav}.

The second of the two tables contains the results. Each row represents an irreducible component of the vacuum manifold. Starting with the first row, the entry \text{$\{ \}$} in the first column
indicates that this component satisfies $F_t\equiv{\rm Re}(F_T)=0$ and $F_\tau\equiv{\rm Im}(F_T)=0$ (in notation we will use throughout) and is therefore a supersymmetric
vacuum. The field values for this vacuum are obtained by setting the polynomials in the column entitled ``\textcolor{MathRed}{0-dim. ideals}" to zero, so for the first row they are found by solving $ct^2-bt-3a=0$ and $\tau =0$.

Similarly, the second row corresponds to another supersymmetric vacuum. The first entry in the last row, $\{ t \}$, indicates that this extremum of the potential has $F_t\neq 0$ and all other F-terms zero (here, simply $F_\tau=0$) and, hence, breaks
supersymmetry in the $t$ direction. {\sf SimpleMinimise} only provides isolated turning points of the potential in field space and so all three components correspond to isolated extrema (as opposed to vacua with flat directions). The absence of a row starting with $\{t , \tau \}$ shows that this model has no vacuum where supersymmetry is broken in the directions of both $t$ and $\tau$.

If no warning messages are given by the package, as is the case in this example, then {\sf SimpleMinimise} returns {\it all} of the isolated turning points of the potential which are present for generic parameter values. Vacua which appear only for non-generic values of the parameters will be discussed in section \ref{constraintssection}.

\subsection{A little more detail: sugra models in \sv}

For more serious calculations {\sf SimpleMinimise} is not the most flexible or convenient way in which to perform such a decomposition of the vacuum space in \sv. Let us repeat the above calculation, therefore, using some mainstream package modules.

In this approach, we first need to define our supergravity model within
\math. This is easily done using the module {\sf CreateModel}. For the above case the
appropriate command reads
\begin{Verbatim}
SingleTmodel = CreateModel[{T},{t},{\[Tau]},-3 Log[T + Conjugate[T]],
  a + b T + c T^2, Par->{a,b,c}];
\end{Verbatim}
In general, this module has five obligatory arguments which are, in order, the list of complex fields, the list of the real parts of these fields, the list of imaginary parts, the K\"ahler potential and
the superpotential. In addition, we have used the option {\sf Par} to
tell the package about the model parameters $a$, $b$, $c$. The output
of {\sf CreateModel} is a {\it model list} of \math\ assignments, here called {\sf SingleTmodel}, of the form {\sf keyword
  $\to$ value}. Each keyword in the model list contains a piece of information about the
supergravity model. Examples of model keywords include {\sf Fields} for the list of complex
fields, {\sf KPot} for the K\"ahler potential, {\sf SPot} for the
superpotential and {\sf Pot} for the scalar potential. For example,
the potential for the above model can be obtained by evaluating \cbox{\texttt{Pot/.SingleTmodel}} which returns
\begin{equation}
	-\frac{2b^2t + 3a(b+2ct) + 2c^2t (t^2 + \tau^2) +bc (5t^2 + 3\tau^2)}{6 t^2}\; .
\end{equation}
This is indeed the bosonic potential of the model \eref{model1}, calculated using \eref{sugrav}. After evaluating {\sf CreateModel} the model list contains all of the  information about the supergravity model which is relevant for an examination of its vacuum structure.

Hidden from the user, the package uses techniques from algorithmic algebraic geometry in order to decompose the vacuum space and describe its properties \cite{Gray:2006gn,Gray:2007yq}. This requires that we process the equations of the supergravity model into a particular form. The required operations are performed by the \math\ module {\sf CalcModel}.
\begin{Verbatim}
SingleTmodel = CalcModel[SingleTmodel];
\end{Verbatim}
After executing {\sf CreateModel} and {\sf CalcModel} a model list contains over $20$ pieces of data, each
referred to by its keyword, and has essentially all of the information one would want for a further analysis of the model. A complete list of the physical content of a model list is given in Appendix~\ref{app:gloss}. 

\begin{figure}[ht]\centering\leavevmode\epsfysize=2cm 
\epsfbox{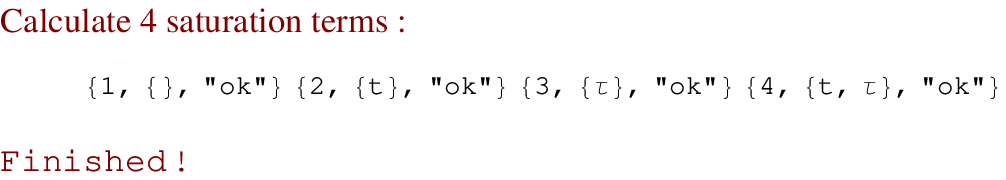}\\
\caption{An example of a \sv\ protocol notebook.}
\label{figure2}
\end{figure}

\begin{figure}[ht]\centering\leavevmode\epsfysize=2cm 
\epsfbox{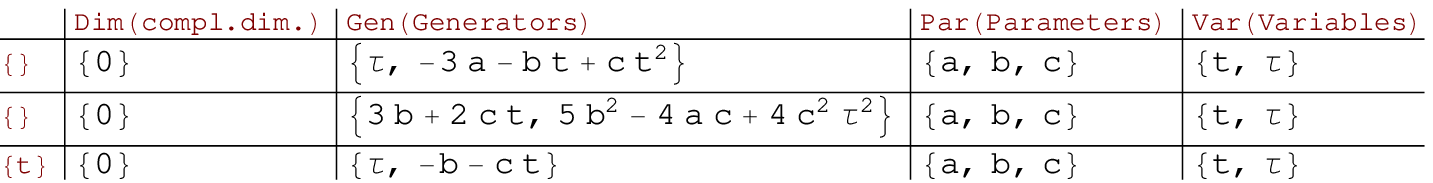}
\caption{An example of the output of {\sf SaturationExpand}.}
\label{figure3}
\end{figure}

As mentioned earlier, one of the key applications of \sv\ is to break up the equations describing the vacua of a model into smaller pieces -- giving one system of equations for each type of vacuum. Let us see how such an analysis is performed given the model list we have just created. We simply evaluate
\begin{Verbatim}
satex = SaturationExpand[SingleTmodel]
\end{Verbatim}
Once this command has been executed information will start to scroll past in the \sing\ window. While this happens a new \math\ protocol notebook appears. Within this is printed the data shown in figure \ref{figure2}.
The terms in the second line of this notebook appear successively until the calculation is finished. Typically after a few seconds, the final result is returned:
\begin{equation}\begin{split}
\big\{ &\{\}\to\big\{\{\text{Gen}\to \{\tau ,-3 a-b t+c t^2\},\text{Dim}\to \{0\},\text{Var}\to\{t,\tau \},\text{Par}\to\{a,b,c\}\big\},\\
&\{\text{Gen}\to  \{3 b+2 c t, 5 b^2-4 a c+4 c^2 \tau ^2 \},\text{Dim}\to \{0\},\text{Var}\to \{t,\tau \},\text{Par}\to \{a,b,c\}\big\}\big\},\\
&\{t\}\to \{\{\text{Gen}\to \{\tau ,-b-c t\},\text{Dim}\to \{0\},\text{Var}\to \{t,\tau \},\text{Par}\to \{a,b,c\}\} \big\}\;.
\end{split}\end{equation}
Each entry in this list represents an irreducible part of the vacuum manifold of our potential, the package has split up the equations for an extremum of the scalar potential as desired. It is sometimes useful to represent results, such as the above, in table form and the package provides a specific command, {\sf TF}, for this purpose. Evaluating
\begin{Verbatim}
satex//TF
\end{Verbatim}
we obtain the table shown in figure \ref{figure3}.
The results in figure \ref{figure3} should be compared to those in figure \ref{figure1} which were obtained by applying {\sf SimpleMinimise} to the same supergravity theory. We see that they are the same and that the presentation is similar in the two cases. In figure \ref{figure3} we may obtain the equations for each of the vacua by setting the polynomials in the column \textcolor{MathRed}{Gen(Generators)} to zero. The first column tells us which F-terms are zero in each vacuum. Thus, as before, we see that this model has two sets of supersymmetric vacua and one non-supersymmetric vacuum with $F_t \neq 0$. As with {\sf SimpleMinimise}, if no warning messages are given by the package, {\sf SaturationExpand} returns {\it all} of the isolated turning points of the potential which appear for generic parameter values. In summary, we have found the complete isolated vacuum structure of the model~\eqref{model1} by performing three simple commands, namely {\sf CreateModel}, {\sf CalcModel} and {\sf SaturationExpand}.

These are, in fact, the three commands being performed during a call to {\sf SimpleMinimise}. However, as we introduce more of the package's abilities it will be useful to have all the information we need close to hand -- this information is stored in the model lists generated by {\sf CalcModel} and {\sf CreateModel}.

\section{Filtering and finding properties of vacua}

In the cases discussed above, the equations returned by {\sf SimpleMinimise} and {\sf SaturationExpand} were so simple that they may be solved trivially to find the vacua. Any physical property of those vacua may subsequently be calculated using the field values obtained. It may not always be the case that these modules simplify the equations to such an extent. We would like, therefore, to have an automated method for extracting the physical properties of the vacua which we discover, which does not require solving the returned equations. \sv\ provides two types of facilities towards this end, {\it filtering} and {\it Sturm queries}. 

\subsection{Filtering}

As \sv\ performs a computation it can discard vacua which do not meet criteria specified by the user. This process, which we call filtering, can greatly speed up computations. It avoids costly calculations being carried out on parts of the problem which are of no physical interest. Many of the modules of \sv\ have a range of optional arguments, some of which deal with the task of filtering\footnote{We will encounter some of the optional arguments found in \sv\ modules in the following sections. A comprehensive list of options for any module can be found in its \math\ help file.}. When omitted in module calls, the filtering options are set to sensible default values.
 
\subsubsection*{Dimension filtering}
 
Perhaps the simplest type of filtering offered by the package is dimension filtering. By specifying an optional argument of the form {\sf Dim}$\rightarrow${\sf list of dimensions}, the user can choose to filter out any vacua which do not have a particular dimension (the number of flat directions). In fact, we have already used this type of filtering in the previous section! In {\sf SimpleMinimise} and {\sf SaturationExpand} this option is set by default to {\sf Dim}$\rightarrow${\sf 0}. Thus, in the previous section, \sv\ was only keeping vacua which are dimension zero -- that is, vacua which do not have flat directions. Had we wished to search for vacua with 1 or 2 flat directions instead we could have evaluated the following.
\begin{Verbatim}
SaturationExpand[SingleTModel, Dim->{1,2}]
\end{Verbatim}
The result, $\{\}$, tells us that there are no vacua with 1 or 2 flat directions in the model defined in \eqref{model1} for generic values of the parameters.

\subsubsection*{Field value filtering}

Another example of physically uninteresting vacua which one may want to filter out are those which lie in an unacceptable region of field space. For example, in a four dimensional theory obtained by compactifying a higher dimensional supergravity, we would wish for the size of the hidden dimensions to be larger than the string scale.
If a particular field were to represent the volume of the hidden dimensions, and if our conventions were such that the string scale was set to $1$, we may therefore only be interested in vacua where this field takes a value greater than $1$.

We can instruct modules, such as {\sf SaturationExpand}, to throw away all vacua which do not satisfy {\it any} set of conditions, as long as those conditions can each be stated as a sign condition on a polynomial in the fields. Let us see how this would work, using the condition described above, in the following model.
\begin{equation}\begin{split}
\label{model2}
K &= -3 \log ( T+ \bar{T}) - \log ( S+\bar{S} ) \, , \quad W = a S + b S T + c T^2\;, \\
T &= t+i\tau,\quad S=s+i\sigma\;.
\end{split}\end{equation}
We begin by creating a model list, containing all of the relevant data of the supergravity theory, and add to this the ancillary mathematical structures the package requires. This is achieved, as in section \ref{start}, by use of the {\sf CreateModel} followed by the {\sf CalcModel} modules.
\begin{Verbatim}
modelnew=CreateModel[{T,S}, {t,s}, {\[Tau],\[Sigma]}, -3 Log[T + Conjugate[T]]
  -Log[S + Conjugate[S]], a S + b S T + c T^2, Par->{a,b,c}];
modelnew=CalcModel[modelnew];	
\end{Verbatim}
We now have a complete model list for the case in hand: {\sf modelnew}. We now wish to apply {\sf SaturationExpand} to this model, filtering out all vacua which do not obey the condition $t > 1$. The relevant command is
\begin{Verbatim}
satexmodelnew = SaturationExpand[modelnew, VarPar->{a->1, b->-1, c->1},
   RRoots->"Probab", NRoots->1, SturmPoly->{t-1}, SturmSigns->1];
\end{Verbatim}
This module call makes use of some new optional arguments\footnote{To find out about the options available for {\sf SaturationExpand} evaluate \texttt{?SaturationExpand}. This returns a brief summary of the module ending with ``{\sf More...}". Click on ``{\sf More...}" to open the \math\ help browser at the relevant page.}. The argument 
\begin{equation}
\msf{VarPar $\bs{\to}$\{a $\bs{\to}$ 1,b $\bs{\to}$ -1,c $\bs{\to}$1\}}
\end{equation}
sets the parameters of the model \eqref{model2} to the indicated values for the duration of this module call. Filtering can only be performed in cases where the parameters have been fixed; whether or not the required filtering conditions are met would otherwise be parameter dependent and would not have a unique answer. Ranges of parameter values can be scanned simply by writing an appropriate loop over module calls in \math. The remaining options are
\begin{equation}
\msf{RRoots $\bs{\to}$ ``Probab", NRoots$\bs{\to}$1, SturmPoly$\bs{\to}$\{t-1\}, SturmSigns$\bs{\to}$1
}
\end{equation}
The first two of these, in order, turn on field value filtering and only keep equations that have at least one real solution. The second two specify $t-1$ as the polynomial whose sign is to be checked and state that the sign of this polynomial should be positive in any vacua to be kept in the output. In short, these options tell the module to only keep vacua for which the polynomial $t-1$ takes positive values, i.e. for which $t>1$
 
It is worth mentioning here that the calls to \sing\ will, of course, take different amounts of time on different computers. On a modern computer the above call will finish in under 30 seconds, however on older machines it may take longer. In the Unix release there is a default maximum time for \sing\ calculations, set at 60 seconds. This is controlled by the option {\sf SingularMaxTime}. For example, to set the maximum time to 120 seconds for a particular calculation, simply add to any call the option\footnote{To change the {\sf SingularMaxTime} setting globally, refer to \texttt{?StringvacuaOptions}} \cbox{\texttt{SingularMaxTime->120}}. 

\begin{figure}[ht]\centering\leavevmode\epsfysize=2.3cm 
\epsfbox{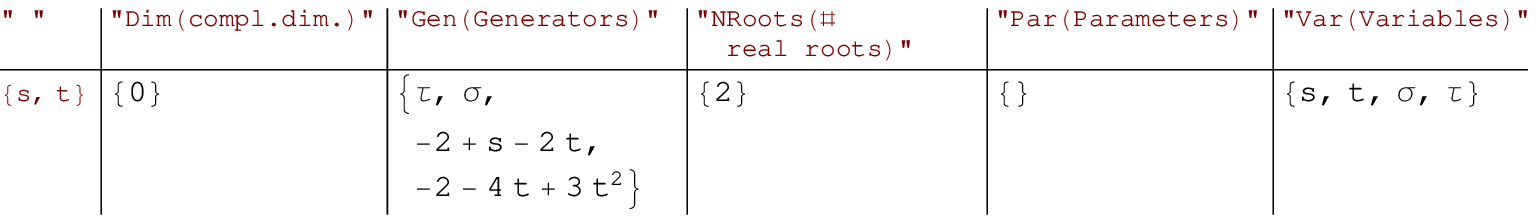}\\
\caption{A filtered output from {\sf SaturationExpand}.}
\label{figure5}
\end{figure}

The result of the above call to {\sf SaturationExpand}, put in table form using \cbox{\texttt{satexmodelnew//TF}}, is as shown in figure \ref{figure5}.

Thus only a single set of vacua exist for this model with $t>1$ and they spontaneously break supersymmetry. The first entry in the results table, \textcolor{MathRed}{\msfb \{s, t\} }, indicates that the real parts of both F-terms are non-zero (while the imaginary parts vanish). The second entry tells us that this vacuum has no flat directions (recall the dimension filtering is set to only keep isolated extrema by default). The third entry is the polynomials which we set to zero to obtain the equations describing the vacua. The fourth column tells us that these equations have $2$ real solutions. As always, since no warning messages were given during this computation, the results in figure \ref{figure5} constitute {\it all} of the vacua of the system which satisfy our filtering constraints.

\subsection{Finding properties of vacua}
Once we have found vacua we will wish to investigate their physical properties (other than those determined by the filtering - which we already know!). This can be achieved in an automated manner using the {\sf SturmQuery} module.

As an example let us examine further the result we obtained at the end of the previous subsection. What else might we like to know about the vacua we have found? We know, from using the {\sf SaturationExpand} options, that the extrema lie in a sensible part of field space. However, we do not yet know if they are stable -- have we found saddle points, minima, or maxima?

To answer this question one would usually compute the eigenvalues of the Hesse matrix in the vacuum and check to see if they are positive, negative or zero -- but this is just a sign test of the type described above! Below, we will show how to formulate the relevant test by hand, in the interest of introducing more of the package modules. However, as stability is such an important issue, the package admits an option `{\sf TestStability}' for {\sf SaturationExpand}, which automates the relevant sign tests. To illustrate, we will repeat the above {\sf SaturationExpand} call, but adding this new option,
\begin{Verbatim}
SaturationExpand[modelnew, VarPar->{a->1, b->-1, c->1},
  RRoots->"Probab", NRoots->1, SturmPoly->{t-1}, SturmSigns->1,
  TestStability->"Hesse"]//TF
\end{Verbatim}
These options, as before, instruct the package to retain only equations which have at least one real solution in the allowed region of field space. The additional, final, option {\sf TestStability} has been set to {\sf "Hesse"} -- this instructs the package to determine the stability of the extrema described by the retained equations. The output, which we have here passed straight to table form, is just as in figure \ref{figure5} (so that we recover the solutions we found previously) except that an additional column has been added to the table, containing
\begin{center}
\begin{tabular}{c|l|c}
{}& {\sf \textcolor{MathRed}{Sturm(Sturm Query)}} & {} \\ \hline
{}& {\sf \{\{Hessian\}, \{\{1,-1\},\{2,2\}\},\{\{1,-1\},\{2,2\}\} \}}&{}
\end{tabular}
\end{center}
How are we to interpret this? Each entry in the column is a list (one for each set of equations retained). The first entry in a list, {\sf \{Hesse\}}, simply labels the type of sign test performed -- here it reminds us that we are testing the signs of the Hesse matrix eigenvalues. There are {\it two} further entries (each a list itself), because the equations have {\it two} real solutions -- in general, then, the list would contain a label and one further list for each real solution. These entries tell us about the signs of the eigenvalues at each of root of the system. In this case they are both identical,
\begin{equation}\label{roots}
	\{\{1,-1\},\{2,2\}\}\;.
\end{equation}
The interpretation of this is simple: at both roots, the extrema have {\it two positive} (the first 2 goes with the +1) and {\it two negative} eigenvalues (the second 2 goes with the -1). Thus the non-supersymmetric vacua we have been studying are actually saddle points.

Let us take a closer look at how the package arrived at this result. The output of the filtered {\sf SaturationExpand} call which led to this result was called {\sf satexmodelnew}. The non-supersymmetric vacuum we found may be extracted from this list by using its label, as given in the first column of figure \ref{figure5}, as follows, \cbox{\texttt{\{s,t\}/.satexmodelnew}} which returns
\begin{equation}\begin{split}\label{eg2vac}
  \{\{\text{Gen}&\to \left\{\tau ,\sigma ,-2+s-2 t,-2-4 t+3 t^2\right\},\text{Dim}\to \{0\}, \text{NRoots}\to \{2\},\\
  &\text{Var}\to \{s,t,\sigma ,\tau \},\text{Par}\to\{\}\}\}
\end{split}\end{equation}
This list contains the information in the table of figure \ref{figure5}, in a format which the modules of \sv\ can understand. We will now pass this information to the {\sf SturmQuery} module to find out about the stability properties of these non-supersymmetric vacua using the sign tests introduced above.

In fact, using {\sf SturmQuery} we can check the signs taken by any rational functions, the signs taken by the determinants of any matrices of rational functions, and the signs taken by the eigenvalues of any matrix with real eigenvalues in the vacua we have found. In our case we of course wish to check the eigenvalues of the Hesse matrix. The first piece of information we require is the Hesse matrix itself. Fortunately this has already been calculated by the package and stored in the model list! The relevant keyword to access this matrix, as may be discovered from the help files, is {\sf Hesse}. Thus evaluating \cbox{\texttt{Hesse/.modelnew}} will return the Hesse matrix of the system. The result is rather lengthy and so we do not reproduce it here.

\begin{figure}[ht]\centering\leavevmode\epsfysize=3cm 
\epsfbox{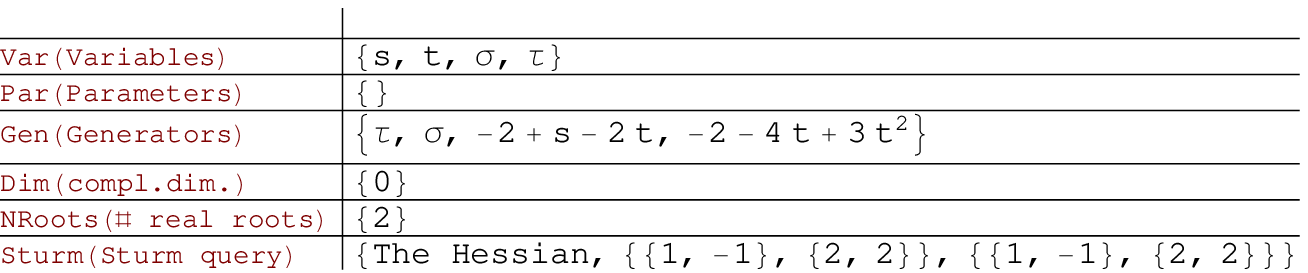}\\
\caption{A example output from {\sf SturmQuery}.}
\label{figure6}
\end{figure}

To identify the signs of the eigenvalues of this matrix in the vacua given by \eqref{eg2vac} we simply perform the following {\sf SturmQuery} call,
\begin{Verbatim}
SturmQuery[{s,t}/.satexmodelnew, TestMatEigen->Hesse/.modelnew,
  VarPar->{a->1, b->-1, c->1}, SturmLabel->"The Hessian", RRoots->"Probab"]
\end{Verbatim}
The first argument above is the component of the vacuum space which we wish to test, as detailed in \eqref{eg2vac}. The second argument is the matrix whose eigenvalues are to be checked, as we just described. The third argument sets the parameters in the Hesse matrix to the correct values -- just those that were selected in the {\sf SaturationExpand} call of the previous section, from which \eqref{eg2vac} was obtained. The fourth argument allows a user-defined label for the results of {\sf SturmQuery}. Finally the fifth argument tells the module which method to use in its computations. The result of this call is shown in figure \ref{figure6} (as earlier we have put the result in table form by appending \texttt{//TF} to the end of the result of the module call).
We see that there is one additional piece of data in this table now, as compared to figure \ref{figure5}: the result of the sign check we have requested. This is precisely the same output we found using {\sf SaturationExpand[... TestStability$\bs\to$ "Hesse"]}, thus we recover the fact that these non-supersymmetric vacua are saddle points. A very large range of properties of vacua can be studied using {\sf SturmQuery}. For more examples we refer the reader to the module's \math\ help file.



\section{Constraints on parameters}
\label{constraintssection}
As well as finding the vacua of a given ${\cal N}=1$ theory, our methods may be used to  identify constraints on the parameters of a model which are necessary for vacua of a given type to exist. As before we shall illustrate this with a simple example. 
Consider the following theory, obtained in \cite{Shelton:2005cf} in the context of non-geometric compactifications of type II string theory.
\begin{equation}\begin{split}
	K &=-3\log(-i(T-\Tb)-\log(-i(S-\Sb))-3\log(-i(U-\overline{U})),\\
	W &= a_0 -3a_1 T +3a_2 T^2-a_3 T^3 \\
	&\hspace{25pt}+ S(-b_0+3b_1 T -3b_2 T^2+b_3 T^3) \\
	&\hspace{25pt}+3U (c_0+(\hat{c}_1+\check{c}_1-\tilde{c}_1)T -(\hat{c}_2+\check{c}_2-\tilde{c}_2)T^2-c_3 T^3)\;,
\end{split}\end{equation}
Here $S,T$ and $U$ are superfields while all other quantities are parameters. The parameters obey the following set of conditions, which come from  consideration of tadpoles and integrability of Bianchi identities. We have,
\begin{equation}\begin{split}
	0&=a_0 b_3 - 3 a_1 b_2 + 3 a_2 b_1 - a_3 b_0 - 16\;,\\
    	0&=a_0 c_3 + a_1 ( \check{c}_2 + \hat{c}_2 - \tilde{c}_2) - a_2 ( \check{c}_1 + \hat{c}_1 - \tilde{c}_1) - a_3 c_0\;,
\end{split}\end{equation}    	
together with the equations below and another set given by exchanging all hats and checks,
\begin{align*}
    0&=c_0 b_2 - \tilde{c}_1 b_1 + \hat{c}_1 b_1 - \check{c}_2 b_0\;,\quad 0=c_0 \tilde{c}_2 - \check{c}_1^2 + \tilde{c}_1 \hat{c}_1 - \hat{c}_2 c_0\;,\\
    0&=\check{c}_1 b3 - \hat{c}_2 b_2 + \tilde{c}_2 b_2 - c_3 b_1\;,\quad 0=c_3 \tilde{c}_1 - \check{c}_2^2 + \tilde{c}_2 \hat{c}_2 - \hat{c}_1 c_3\;,\\
    0&=c_0 b3 - \tilde{c}_1 b_2 + \hat{c}_1 b_2 - \check{c}_2 b_1\;,\quad 0=c_3 c_0 - \check{c}_2 \hat{c}_1 + \tilde{c}_2 \check{c}_1 - \hat{c}_1 \tilde{c}_2\;,\\
    0&=\check{c}_1 b_2 - \hat{c}_2 b_1 + \tilde{c}_2 b_1 - c_3 b_0\;,\quad 0=\hat{c}_2 \tilde{c}_1 - \tilde{c}_1 \check{c}_2 + \check{c}_1 \hat{c}_2 - c_0 c_3\;.
\end{align*}
As an example, we shall describe how to find further constraints on the parameters of this system which are necessary (and in this case sufficient) for the existence of supersymmetric Minkowski vacua. These vacua are obtained as solutions to the equations,
\begin{equation}
\label{minkeqns}
\partial_T W = 0 \;, \; \partial_S W =0 \;, \; \partial_U W =0 \;, \; W=0 \;,
\end{equation}
together, of course with the vanishing of the conditions on the parameters given above. This large and complicated set of equations contains information about what values the flux parameters can take in order for solutions to exist, but this data is entangled with all of the other information about the extrema (such as field values). Our objective is to separate out the information concerning the constraints on the parameters.
This may be achieved using the package routine {\sf Elimination}. 

\begin{figure}[ht]\centering\leavevmode\epsfysize=8cm 
\epsfbox{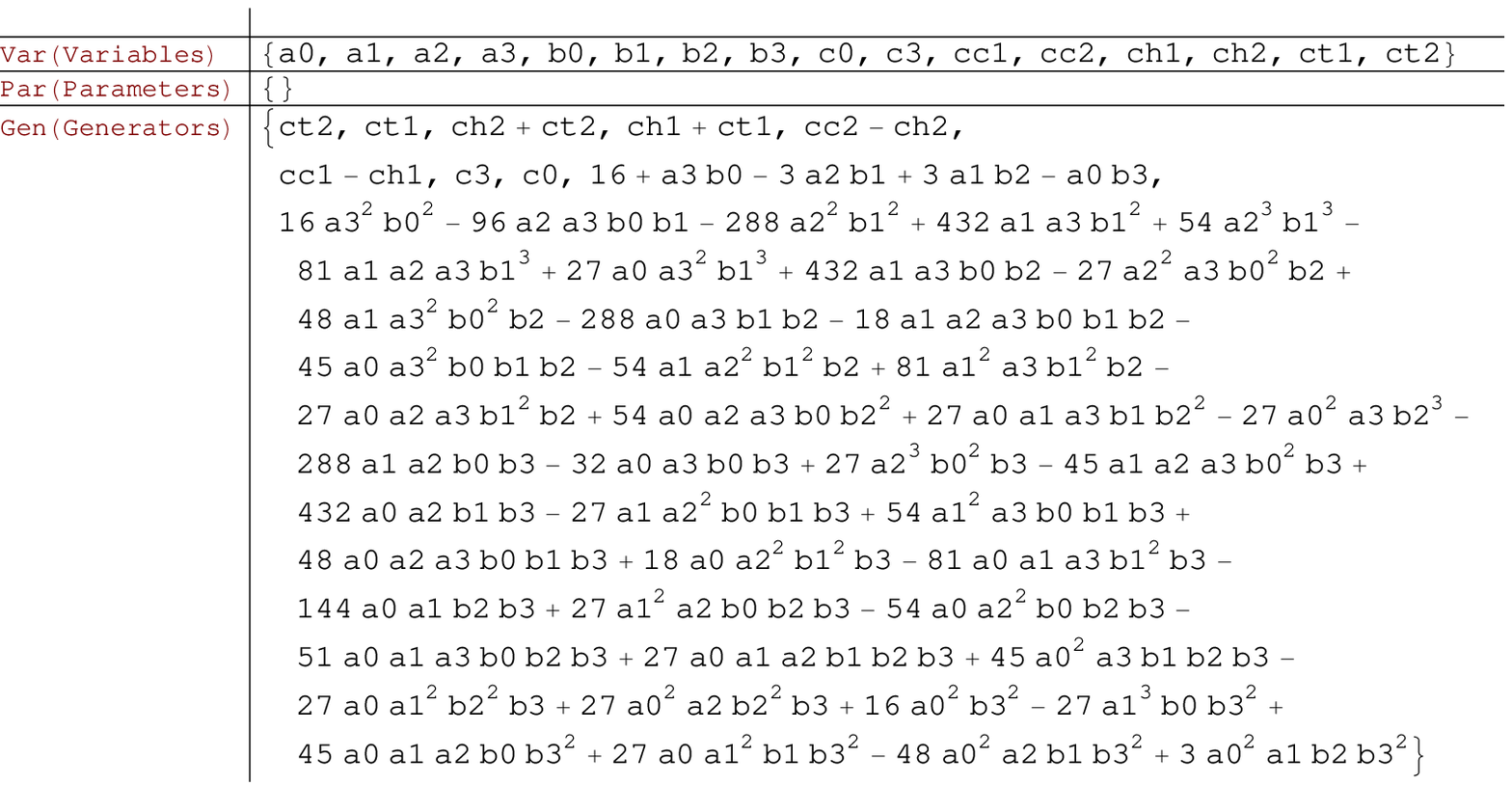}\\
\caption{Constraints on parameters obtained using {\sf Elimination}.}
\label{figure8}
\end{figure}

First, we must collect the various equations describing the vacua we wish to analyse. We create the model list for the above system using the commands {\sf CreateModel} and {\sf CalcModel} as in earlier sections. Let us call the resulting model list {\sf model}. The equations in \eqref{minkeqns}, being information about our supergravity model, are now stored in this model list and may be extracted, in a form \sv\ modules will understand, by evaluating the following command.
\begin{Verbatim}
{IDSPot/.model, ISPot/.model}
\end{Verbatim}
See the appendix for a description of these keywords. Next we must enter the conditions on the parameters given above into \math. Using the usual \math\ syntax, these are included as a list of the polynomials which the conditions dictate vanish, let us call it {\sf Constraints}.
\begin{Verbatim}
Constraints={a0 b3 - 3 a1 b2 + 3 a2 b1 - a3 b0 - 16,...}
\end{Verbatim}
Now we have entered all of the relevant equations we may simply plug them into the module {\sf Elimination} to obtain our constraints.  
The first argument in this module call is simply the collection of equations we wish to vanish. The optional argument list {\sf VarPar} instructs the module to eliminate some set of variables from the equations it is passed. Here, we set \texttt{VarPar->\{T,S,U\}} in order to eliminate the field variables $T$, $S$ and $U$ from the equations. A final argument {\sf Meth} is optional. Here we use it to select one of the faster methods of elimination available. The full command to be entered is
\begin{Verbatim}
NewConstraints = Elimination[{IDSPot/.model, ISPot/.model, Constraints},
  VarPar->{T,S,U}, Meth->"Prod"];
\end{Verbatim}
We are left with {\sf NewConstraints} which contains a list of polynomials. As usual we may display this object in a nice form by evaluating \cbox{\texttt{NewConstraints//TF}}.
The result is shown in figure \ref{figure8}.
When set to zero, the polynomials in the box ``\textcolor{MathRed}{Gen (Generators)}" represent the constraints on the flux parameters which are necessary and sufficient for the existence of supersymmetric Minkowski vacua\footnote{Often, one can apply one of the other \sv\ modules, {\sf PrimDec}, to the output of {\sf Elimination} in order to separate this list into smaller, simpler subsets, the vanishing of each subset representing a different way to solve the full set of constraints.}. Note that in figure \ref{figure8} we have relabelled the parameters in an obvious way for ease of input into \math\ (ct1 for $\tilde{c}_1$, etc).

Use of constraints, such as those found in this section, is one of the ways in which vacua which only appear for non-generic values of the parameters can be found using \sv. One can simply solve for a sufficiently large number of parameters until the constraints for the required type of vacuum are automatically satisfied. Alternatively if one is running loops over calls to {\sf SaturationExpand}, one for each set of flux parameters, the constraints in this section can be used to greatly speed up the calculation. Any set of parameters which do not satisfy the constraints need not be investigated.

\section{Non-perturbative effects}

So far we have addressed models for which the scalar potential, the F-terms, and all other relevant quantities are all rational functions of the fields. In many cases, such as in models where non-perturbative contributions to the superpotential play an important role, we may be interested in more complicated situations. These can be addressed using the \sv\ package by the simple expedient of replacing any non-polynomial terms in the equations with dummy variables. As in earlier sections we will illustrate this with some examples.

\subsection{Creating non-perturbative models}
Consider the following toy model as an example.
\begin{equation}\label{system}\begin{split}
	W &= T + \exp(-S)\; ,\quad K =-\log(T+\Tb)-3\log(S+\overline{S})\;, \\
	T &= t +i\tau\; ,\quad S=s+i\sigma\;
\end{split}\end{equation}
The supergravity model is set up in the package using {\sf CreateModel}, just as before.
\begin{Verbatim}
model = CreateModel[{T,S}, {t,s}, {\[Tau],\[Sigma]},
  -Log[T + Conjugate[T]] - 3 Log[S + Conjugate[S]], T + Exp[-S]];
\end{Verbatim}
Before we can apply our methods we must find a way to transform the system into a set of equations \sv\ can understand -- essentially into polynomials. This may be achieved by using dummy variables to represent non-algebraic objects. A good choice of dummy variables, in this case, is the set,
\begin{equation}\label{choice}
	X\equiv \exp(-S),\quad Y\equiv\exp(-\overline{S})\;, \quad x\equiv \exp(-s),\quad y\equiv\exp(-i\sigma)\; .
\end{equation}
These variables are built into the model as a set of replacements included in {\sf CalcModel} (which, recall, takes supergravity objects and converts them into expressions the package can use) through two options. These are {\sf CDummies$\to${...}}, a list of replacements for the superfields, and {\sf Dummies$\to${...}}, a list of replacements for the real fields. The dummy variables defined in (\ref{choice}) are entered as shown below.
\begin{Verbatim}
fullmodel = CalcModel[model, CDummies->{Exp[-S]->X, Exp[-Conjugate[S]]->Y},
  Dummies->{Exp[-s]->x, Exp[-I\[Sigma]]->y}];
\end{Verbatim}
To see the effect of the dummy variables, compare the supergravity potential, \begin{Verbatim}
Pot/.fullmodel
\end{Verbatim}
\begin{equation}
	\frac{\mathrm{e}^{-2s}}{48 s^3 t}\bigg( 3+12s + 4s^2 + 3\mathrm{e}^{2s}t^2+4\mathrm{e}^{2s}\tau^2+6\mathrm{e}^s(2s-1)t\cos(\sigma)-6\mathrm{e}^s(2a+1)\tau\sin(\sigma)  \bigg)\;,
\end{equation}
with its numerator, converted into a polynomial,
\begin{Verbatim}
IPot/.fullmodel
\end{Verbatim}
\begin{equation}\begin{split}
	&\{\text{Var} \to \{t, s,  \tau,  \sigma, x, y\}, \text{Par} \to \{\}, 
    \text{Gen} \to \{ -3tx + 6  s  t  x + 3  t ^2  y + 3  x ^2  y + 12  s  x ^2  y + 4  s ^2  x ^2  y \\
    &- 3  t  x  y ^2 + 6  s  t  x  y ^2 + 3   i  x   \tau + 6   i  s  x   \tau - 3   i  x  y ^2   \tau - 6   i  s  x  y ^2   \tau + 3  y   \tau ^2\}\} \;.
\end{split}\end{equation}
Here, the dummy variables $\{x,y\}$ have been appended to the list of true variables $\{t,s,\tau,\sigma\}$ and the potential has been reduced to a polynomial as desired.
\subsection{Finding vacua}
The routine \texttt{SaturationExpand} may be used to identify and classify all vacua of this model in a few seconds. Note that we are working in an enlarged field space with dummy variables, in which true isolated vacua may appear to have flat directions. To ensure that this is accounted for, we adjust the {\sf Dim} option in {\sf SaturationExpand} to retain vacua with any number of flat directions up to and including the number of (real field) dummy variables included. {\sf SaturationExpand} is then called using
\begin{Verbatim}
satexNP = SaturationExpand[fullmodel, Dim ->{0,1,2}]	
\end{Verbatim}
which returns a list {\sf satexNP} of vacua classified by their SUSY breaking in a manner which is now familiar (see, for example, figure \ref{figure5}). Consider the vacuum space in which the imaginary parts of both F-terms vanish but the real parts are non-zero. We can extract the vacua of this kind from the full list by using their identifier, thus \cbox{\texttt{\{s,t\}/.satexNP}}.
This list contains two potential vacua, as seen by evaluating \cbox{\texttt{Length[\{s,t\}/.satexNP]}}. Extracting one of these pieces using \cbox{\texttt{(\{s,t\}/.satexNP)[[1]]}} we find
\begin{equation}\begin{split}
    &\{\text{Gen} \to \{ \tau, 1 + y,   -3    t ^2 + 7  t  x + 2  s  t  x + 2  x ^2 + 4  s  x ^2,   -7    t - 2  s  t + x + 8  s  x + 4  s ^2  x,\\
    &-11  t ^2 + 2  s  t ^2 + 11  t  x + 6  x ^2 + 16  s  x ^2,   -15    t - 2  s  t + 2  s ^2  t - 3  x - 4  s  x, \\
    &3  t ^3 - 24  t ^2  x + 23  t  x ^2 + 10  x ^3 + 24  s  x ^3, 3  t ^4 - 18  t ^3  x + 11  t ^2  x ^2 - 28  t  x ^3 - 4  x ^4,\\
    &-18   - 78  s - 41  s ^2 + 4  s ^3 + 4  s ^4\},\text{Dim} \to \{2\}, \text{Var} \to \{s, t, x, y,  \sigma,  \tau\}, 
      \text{Par} \to \{\}\}
\end{split}\end{equation}
Thus, even in these non-perturbative cases, {\sf SaturationExpand} can be used to break up the equations describing the vacuum space into smaller,  more manageable pieces. In the above we could easily solve the equation given by setting the last polynomial in {\sf Gen} to zero to find $s$ and hence $x$. The solutions for $\tau$ and $y$ (and thus $\sigma$) are evident from the first two polynomials. These solutions could then be plugged into the remaining polynomials to find $t$. In other words, in many cases the decomposition of the vacuum space provided by {\sf SaturationExpand} makes the equations to be solved trivial. In situations where this is not the case the algorithm developed in \cite{Gray:2007yq} should be used. This algorithm can be executed using the modules of the package. This is, however, more involved than the level of this simple tutorial and we refer the user to \cite{Gray:2007yq} and to the package help files for more information.

The essential point we wish to convey is that one of the core keys to the power of \sv\ -- breaking up the equations for the vacuum space into smaller pieces -- is applicable even in these non-polynomial cases. 
\subsection{Constraints in non-perturbative models}

The other application of \sv\ which we have introduced in this tutorial, finding constraints on parameters, also works in the case where non-perturbative effects are present.

As an example consider the supergravity model,
\begin{equation}\begin{split}
\label{npconstmodel}
	K &= -3 \log\big(T+\bar{T}\big)-3 \log\big(Z+\bar{Z}\big)-\log\big(S+\bar{S}\big)\;,\\
	W &= i(\xi+i e T)+(\epsilon+i p T) Z + i (\mu+i q T)Z^2/2 + \lambda\, \exp(-c S)\;,\\
	T &= t + i \tau,\ Z = z + i \zeta,\  S = s + i \sigma\;.
\end{split}\end{equation}
which is obtained by compactifying heterotic string theory on a generalised half-flat manifold in the presence of gaugino condensation \cite{Gurrieri:2004dt,deCarlos:2005kh}. We can input this model into \sv\ using {\sf CreateModel} and {\sf CalcModel} as usual, using the {\sf CDummies} and {\sf Dummies} options to specify the replacements for the exponentials. To clarify the presentation we here first define the K\"ahler and superpotentials,
\begin{Verbatim}
K = -3 Log[T+Conjugate[T]] -3 Log [Z+Conjugate[Z]]-Log[S+Conjugate[S]];
W = I(\[Xi]+I e T)+(\[Epsilon]+I p T) Z + I (\[Mu]+I q T) Z^2/2
  +\[Lambda] Exp[-c S];
\end{Verbatim}
These are then entered into {\sf CreateModel} using the identifiers {\sf K} and {\sf W} to create the supergravity model,
\begin{Verbatim}
NPmodel = CreateModel[{T,S,Z}, {t,s,z}, {\[Tau],\[Sigma],\[Zeta]}, K, W,
  Par->{\[Xi], e, \[Epsilon], p, \[Mu], q, \[Lambda], c}];
\end{Verbatim}
Now we use {\sf CalcModel} to create the extended model list, including our choice of dummy variables for the exponential function,
\begin{Verbatim}
NPmodel = CalcModel[NPmodel, CDummies->{Exp[-c S]->X, Exp[-c Conjugate[S]]->Y},
  Dummies->{Exp[-c s]-> x, Exp[-I c \[Sigma]]-> y}];
\end{Verbatim}
There is one condition on the parameters which follows from the integrability of the Bianchi identity, $q\epsilon-p\mu=0$, which we enter simply as
\begin{Verbatim}
BIcondition = q \[Epsilon] - p \[Mu];
\end{Verbatim}
We can find further constraints on the parameters of this model which are necessary for the existence of supersymmetric vacua satisfying certain physical conditions. For various physical reasons we might be interested in vacua with field values $t=1$, $s=24$, and $\sigma=0$, for example. 
We require, then, that the F-terms, and the polynomials $t-1$, $s-24$ and $\sigma$ all vanish in our vacua.

\begin{figure}[ht]\centering\leavevmode\epsfysize=5cm 
\epsfbox{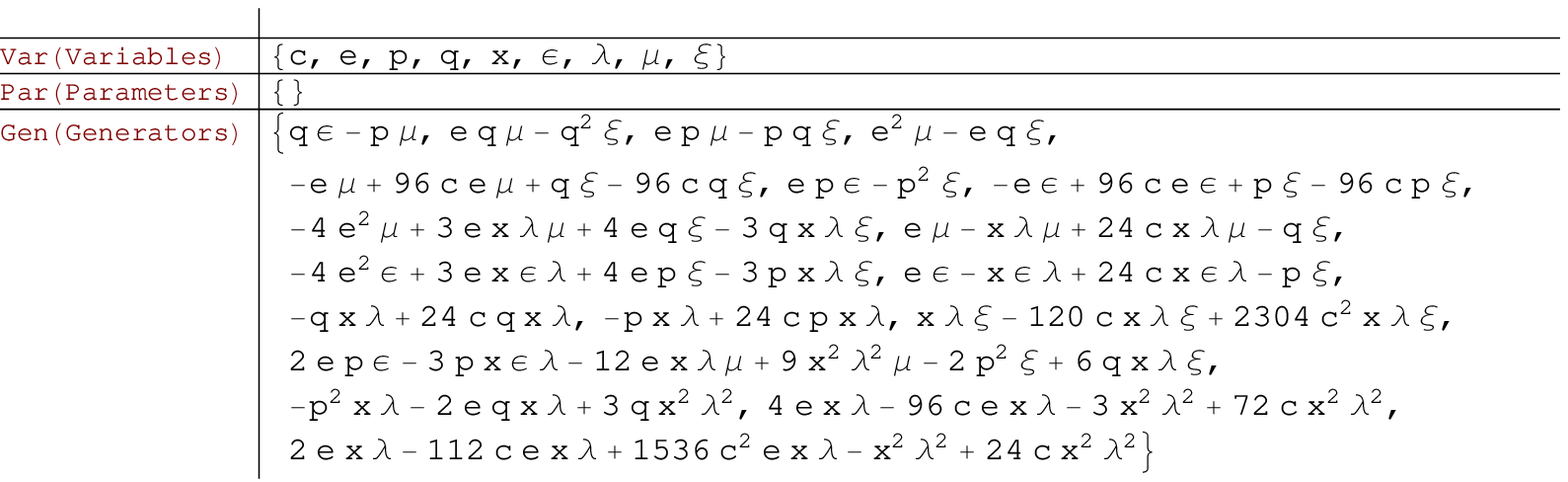}\\
\caption{Constraints in a non-perturbative example using {\sf Elimination}.}
\label{figure7}
\end{figure}

Note that setting the value of $\sigma=0$ implies a fixed value for $y\equiv\exp(-ic\sigma)=1$, so we should also set the polynomial $y-1$ to vanish.
At this stage, even though we have chosen a value for $s$, the value of the dummy variable $x$ still depends on the parameter $c$. As such, we cannot fix its value. To find our constraints, we simply collect all of our equations and then eliminate all of the fields using the command {\sf Elimination}, exactly as in section \ref{constraintssection}.
\begin{Verbatim}
ConstraintsNP = Elimination[{IFTerms/.NPmodel, BIcondition, t-1, s-24,
  \[Sigma], y-1}, VarPar-> {s, t, y, z, \[Zeta], \[Sigma], \[Tau]}];
\end{Verbatim}
The output of this module call is shown in figure \ref{figure7}, put in table form by evaluating\\
\newline
\cbox{\texttt{ConstraintsNP//TF}}.

The returned polynomials in the entry labelled ``\textcolor{MathRed}{Gen (Generators)}" should be set to zero to obtain the desired constraints. The vanishing of these polynomials describes necessary constraints on the flux parameters of this model (necessary and sufficient constraints only being obtained in the case of perturbative supersymmetric Minkowski vacua) for the existence of vacua of the desired type \cite{Gray:2006gn,Gray:2007yq}. As mentioned in section \ref{constraintssection}, constraints such as these can frequently be simplified using other \sv\ modules - for example {\sf PrimDec}. The user is directed to the \math\ help files for further information.

\section{STRINGVACUA: other applications}
Since the package is tailored to the fast manipulations of
systems of polynomials, its applications clearly extend far beyond finding
the vacua of string-inspired or derived models as described above. Problems such as finding
the dimension of the space defined by a system of polynomials, or how many real
roots there are, etc., are ubiquitous in countless fields of modern
research. The convenient \math\ interface we have developed, 
which requires neither an understanding of computational algebraic geometry nor of
the program \sing, should prove useful to researchers in diverse fields
of interest. In this short section we will illustrate the solution of questions one may encounter in more abstract situations.

\begin{figure}[ht]\centering\leavevmode\epsfysize=7cm
\epsfbox{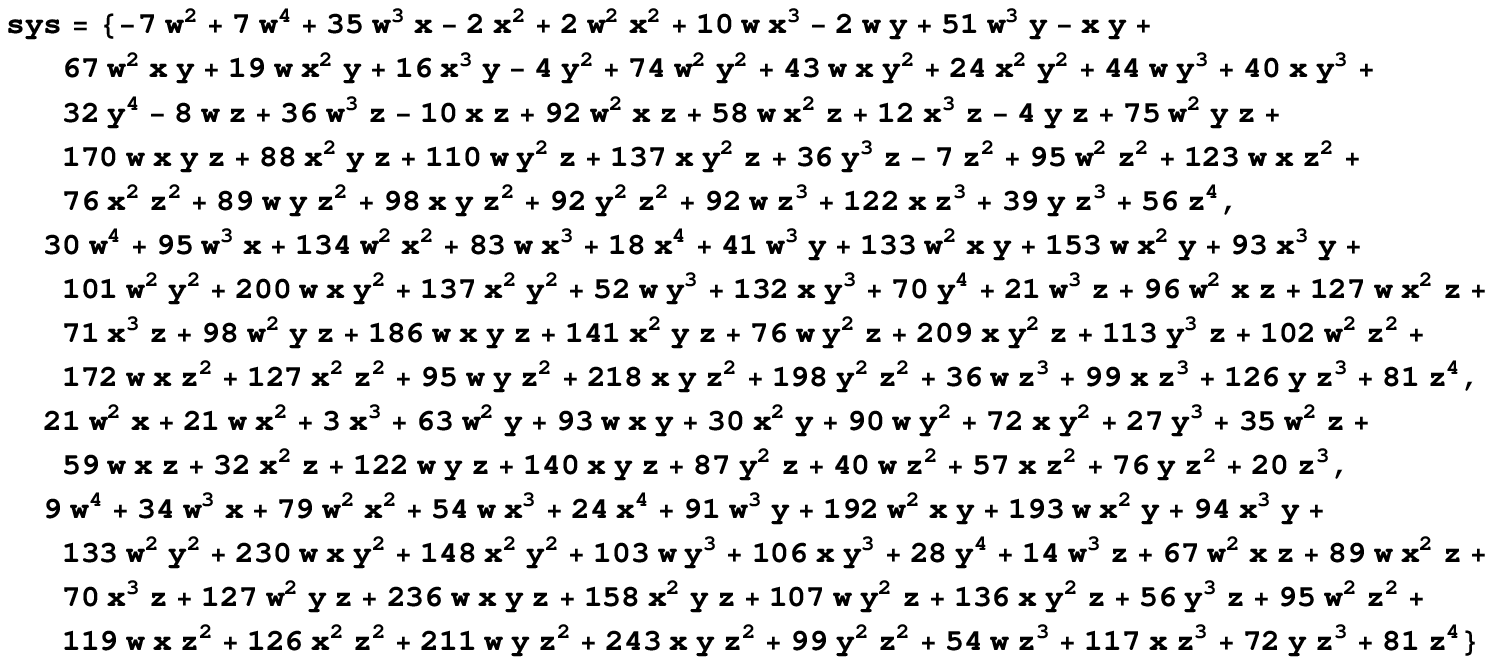}\\
\caption{A given system of polynomial equations.}
\label{f:egpoly}
\end{figure}

Suppose we are given a formidable-looking system of polynomials, given
in figure \ref{f:egpoly} and called {\sf sys}, and we wish to check whether its zero-locus has
dimension zero (i.e., no flat directions) and if so find how many real
roots this system has. A system of polynomials, in the language of
algebraic geometry, is called an {\it ideal}. Ideals are stored within \sv\ as a list of assignments, 
called an {\it ideal list}, as follows.
\begin{verbatim}
ideal = {Gen->{polynomials}, Var->{variables}, Par->{parameters}}
\end{verbatim}
Additional assignments can be added to ideal lists by some of the \sv\ modules -- we shall see an example of this shortly.

Given this general structure, the dimension in our current example is easily calculated using the command
\begin{Verbatim}
Dim/.DimIdeal[{Var-> {x, y, z, w}, Gen-> sys}]
\end{Verbatim}
As described above, the {\sf Var $\rightarrow$} option specifies that the
variables are $\{x,y,z,w\}$, the {\sf Gen $\rightarrow$ sys}
dictates that the system to be studied is our list {\sf sys} of 4
polynomials in 4 variables and there is no {\sf Par} assignment in this case as there are no parameters in our system. 
The command {\sf DimIdeal[~]} then tells us
the dimension of the space thus defined; it adds the assignment {\sf Dim $\rightarrow$ \{the dimension\}}
to the ideal list. Here we have extracted the dimension from this list using {\sf Dim /.} The
answer is very quickly found to be  {\sf {0}}.

We can now try to find the real
roots. The \math\ command {\sf Solve[ ]} is readily seen to be
ineffective in solving this system. However, methods of computational
algebraic geometry can first reduce this system into much simpler pieces
by a technique known as {\it primary decomposition}. What this means is
that the package can split {\sf sys} up into a finite number of simpler
components, the union of whose zero-loci is the zero-locus of {\sf sys}
itself -- in other words, solving any of the simpler sets of equations gives a solution to the full set {\sf sys}:
\begin{Verbatim}
pieces = PrimDec[{Var->{x,y,z,w}, Gen->sys}];
\end{Verbatim}
The output is rather complicated so we will not present it here; suffice
it to say that {\sf pieces} is a list of length 11, check using \cbox{\texttt{Length[pieces]}}, meaning that there are 11 simple components to {\sf sys}, each of dimension 0 and each represented by its own ideal list\footnote{By default, {\sf PrimDec} retains only dimension 0 pieces. As before this may be adjusted using the {\sf Dim$\to$} option.}. Now, we can
find the number of real roots of any 0-dimensional system of polynomials
with the command {\sf NumRoots[~]}. This adds the assignment {\sf NRoots $\rightarrow$ \{number of roots\}} to an ideal list. We can then extract the number with the keyword {\sf NRoots} in the same way as we extracted the dimension above.

We can perform this for each of the 11 components by entering
\begin{Verbatim}
Table[NRoots/.NumRoots[pieces[[i]]], {i, 1, 11}]
\end{Verbatim}
giving us
\[
\{ \{ 1\} ,\{ 0\} ,\{ 0\} ,\{ 1\} ,\{ 0\} ,\{ 0\} ,\{ 1\} ,\{ 0\} ,\{
0\} ,\{ 0\} , \{ 0\} \}
\]
Therefore, we see that eight of the components do not have real roots while
three of them have a single real root. In other words, {\sf sys} has a total of 3 real roots.

We can, of course, do more. Since we have the explicit expressions for
the actual pieces we can find what the roots are explicitly. We see from
the above that the 1st, 4th and 7th components have real roots.
Executing \cbox{\texttt{pieces[[1]]}} and similarly extracting {\sf pieces[[4]]} and {\sf pieces[[7]]} will give us the relevant equations. In fact, here they all turn out to be
\begin{center}
{\sf \{Gen $\rightarrow$ \{z, y, x, w\}, Dim $\rightarrow$ \{0\},
Var $\rightarrow$ \{w, x, y, z\}, Par $\rightarrow$ \{\}\}}
\end{center}
We see that the {\sf Gen} keyword is set to $\{z, y, x,
w\}$. To find the real roots we must therefore solve the equations obtained by setting each of the polynomials in this list to zero - which is just the origin. In other words, all of the roots here are at
$(0,0,0,0)$ and therefore the original system, {\sf sys}, has a triple root at
the origin and no other real roots.
\section{Summary}

We have briefly described, illustrating our discussion with simple examples, how the \math\ package \sv\ may be used to aid in the investigation of vacuum configurations of models of string phenomenology. We have focused on two main applications. First, we described how to split the large equations describing the vacuum space of a model up into smaller systems, each of which can be dealt with more easily. Second, we discussed how to obtain constraints on the parameters in a theory necessary for the existence of given types of vacua.

In fact, in this short note, we have only used 7 of the package's
17 modules and we have far from exhausted the flexibility provided by optional arguments even in those. The message is that there is much more \sv\ can do to aid the user than the simple automated routines described in this document. Many of the methods imported to \math\ from \sing\ by this package are not restricted to use in string theory, but will benefit researchers in any area where a problem involves polynomial computations.  The papers \cite{Gray:2006gn,Gray:2007yq}, along with the \math\ help files for the various modules, should go a long way towards helping the user in apply the techniques of \sv\ in a more flexible manner. This, as always with a computer algebra system, can lead to huge improvements in what may be achieved using the package.

\section*{Acknowledgments}

It is a pleasure to thank Beatriz de Carlos, Vishnu Jejjala, Sven-Ludwig Krippendorf, Eran Palti, Fernando Quevedo, Mario Serna, and George Weatherill for testing and making useful suggestions on the beta version of Stringvacua. J.~G.~is supported by STFC, A.~L.~is supported by the EC 6th Framework Programme MRTN-CT-2004-503369, and Y.-H. H is supported by the FitzJames Fellowship of Merton College, Oxford.

\appendix
%
%
\section{Glossary -- supergravity model list keywords}
\label{app:gloss}

\subsection{CreateModel}
The command \texttt{model=CreateModel[...]} takes a list of fields, a K\"ahler potential and a superpotential, and creates a model list, here called {\sf model}. A model list contains essentially all of the information about a supergravity theory which is useful in an analysis of its vacuum space. Objects, such as the F-terms and the scalar potential, can be extracted via `keywords' by executing a command of the form
\begin{Verbatim}
keyword/.model
\end{Verbatim}
where \texttt{keyword} can be any of the following:\\
\newline
\begin{tabular}{p{1.4in}|p{4.9in}}
{\bf Keyword} & {\bf Description}\\ \hline\hline
\\
{\sf Fields} & A list of the superfields in the model, e.g. $\{T\}$. \\
{\sf ReFields} & A list of the real parts of the superfields in the model, e.g. $\{t\}$.\\
{\sf ImFields} & A list of the imaginary parts of the superfields in the model, e.g. $\{\tau\}$.\\
{\sf ComplexToReal} & A list of the relations between the {\sf Fields} and their real and imaginary parts, in the form $\{T\to t+i\tau\}$.\\
{\sf Par} & A list of the flux parameters in the model, e.g. $\{a\}$.\\
{\sf KPot} & The K\"ahler potential. \\
{\sf SPot} & The superpotential.\\
{\sf DeltaPot} & The non-supersymmetric extension to the scalar potential.\\
{\sf DSPot} & A list of first derivatives of the superpotential with respect to the superfields, in the form ``\{field $\to$ derivative with respect to field\}".\\
{\sf FTerms} & A list of the F-terms, in the form ``\{field $\to$ F-term\}". \\
{\sf Pot} & The supergravity scalar potential, calculated from the superpotential, K\"ahler potential and non-SUSY extension. \\
{\sf DPot} & A list of the first derivatives of the scalar potential, in the same format as {\sf DSPot}, above.\\
{\sf Hesse} & The Hessian matrix of the scalar potential, computed with respect to the real fields.\\\hline\hline
\end{tabular}
\subsection{CalcModel}
{\sf CalcModel} creates an extended model list which contains the keywords above and added assignments where the supergravity objects have been processed into a form suitable for our methods -- an ``{\sf I}" is prefixed to the above keywords to indicate this. For example, {\sf Pot/.model} is now accompanied by {\sf IPot/.model}. All data is accessed in the manner described above. The following optional inputs of {\sf CalcModel} are also keywords: \\
\newline
\begin{tabular}{p{1.4in}|p{4.9in}}
{\bf Keyword} & {\bf Description}\\ \hline\hline\\
{\sf CDummies} & A list of replacement rules for non-algebraic functions of superfields, in the form ``function$\to$ dummy variable".\\
{\sf Dummies} & A list of replacement rules for non-algebraic functions of real fields, in the form ``function$\to$ dummy variable".\\
{\sf RDummies} & A list of rules for replacing non-algebraic functions with dummy variables, in the form ``function$\to$ dummy variable".\\ \hline\hline

\end{tabular}

\end{document}